\documentclass[useAMS,usenatbib]{mn2e}
\usepackage{graphicx}

\newcommand\aj{{AJ}}%
\newcommand\apj{{ApJ}}%
\newcommand\apjl{{ApJ}}%
\newcommand\apjs{{ApJS}}%
\newcommand\mnras{{MNRAS}}%
\newcommand\aap{{A\&A}}%
\newcommand\nat{{Nature}}%

\newcommand{\ltsim}{\raisebox{-.5ex}{$\;\stackrel{<}{\sim}\;$}}
\newcommand{\gtsim}{\raisebox{-.5ex}{$\;\stackrel{>}{\sim}\;$}}
\newcommand{\vFWHM}{\ifmmode V_{\mbox{\tiny FWHM}} \else $V_{\mbox{\tiny FWHM}}$ \fi}
\newcommand{\et}{et al.\ }

\newcommand{\Lop}{\ifmmode L_{5100} \else $L_{5100}$\fi}
\newcommand{\Lthree}{\ifmmode L_{3000} \else $L_{3000}$\fi}
\newcommand{\lledd}{\ifmmode L/L_{\rm Edd} \else $L/L_{\rm Edd}$\fi}
\newcommand{\lbol} {\ifmmode L_{\rm bol} \else $L_{\rm bol}$\fi}
\newcommand{\lamLlam}{\ifmmode \lambda L_{\lambda} \else $\lambda L_{\lambda}$\fi}

\newcommand{\mbh}{\ifmmode M_{\rm BH} \else $M_{\rm BH}$\fi} 
\newcommand{\mbul}{\ifmmode M_{\rm Bulge} \else $M_{\rm bulge}$\fi} 
\newcommand{\mstar}{\ifmmode M_{*} \else $M_{*}$\fi} 
\newcommand{\mhost}{\ifmmode M_{\rm Host} \else $M_{\rm Host}$\fi}
\newcommand{\mm}{\ifmmode M_{*}/M_{\rm BH} \else $M_{*}/M_{\rm BH}$\fi}
\newcommand{\mmwp}{\ifmmode \left(M_{*}/M_{\rm BH}\right) \else $\left(M_{*}/M_{\rm BH}\right)$\fi}
\newcommand{\ml}{\ifmmode M_{*}/L_{*} \else $M_{*}/L_{*}$\fi}
\newcommand{\mlwp}{\ifmmode \left(M_{*}/L\right) \else $\left(M_{*}/L\right)$\fi}
\newcommand{\mlk}{\ifmmode \left(M_{*}/L_{K}\right) \else $\left(M_{*}/L_{K}\right)$\fi}
\newcommand{\sigs}{\ifmmode \sigma_{*} \else $\sigma_{*}$\fi}

\newcommand  \kms      {\ifmmode {\rm km\,s}^{-1} \else km\,s$^{-1}$ \fi}

\newcommand  \ergs     {\ifmmode {\rm ergs\,s}^{-1} \else ergs\,s$^{-1}$ \fi}
\newcommand  \ergcms   {\ifmmode {\rm ergs\,cm}^{-2}\,{\rm s}^{-1} \else ergs\,cm$^{-2}$\,s$^{-1}$\fi}
\newcommand  \ergcmsA {\ifmmode{\rm ergs\,cm}^{-2}\,{\rm s}^{-1}\,{\rm\AA}^{-1} \else ergs\,cm$^{-2}$\,s$^{-1}$\,\AA$^{-1}$\fi}
\newcommand \ergcmsHz {\ifmmode{\rm ergs\,cm}^{-2}\,{\rm s}^{-1}\,{\rm Hz}^{-1} \else ergs\,cm$^{-2}$\,s$^{-1}$\,Hz$^{-1}$\fi}
\newcommand  \phcms    {\ifmmode {\rm ph\,cm}^{-2}\,{\rm s}^{-1} \else ,ph\,cm$^{-2}$\,s$^{-1}$\fi}
\newcommand  \phcmsA   {\ifmmode {\rm ph\,cm}^{-2}\,{\rm s}^{-1}\,{\rm\AA}^{-1} \else ph\,cm$^{-2}$\,s$^{-1}$\,\AA$^{-1}$\fi}

\newcommand \zandz	{z$\simeq$1 and z$\simeq$2}

\newcommand\Msun{\ifmmode M_{\odot} \else $M_{\odot}$\fi}
\newcommand\msun{\ifmmode M_{\odot} \else $M_{\odot}$\fi}
\newcommand\Lsun{\ifmmode L_{\odot} \else $L_{\odot}$\fi}

\newcommand\mpyr{\ifmmode \Msun\,{\rm yr}^{-1} \else $\Msun\,{\rm yr}^{-1}$ \fi}

\newcommand \Hbeta {\ifmmode {\rm H}\beta \else H$\beta$ \fi}
\newcommand \hb    {\ifmmode {\rm H}\beta \else H$\beta$ \fi}
\newcommand  \mgii  {\ifmmode {\rm Mg}{\textsc{ii}} \else Mg\,{\sc ii} \fi} 
\newcommand  \MgII  {\ifmmode {\rm Mg}\,{\sc ii}\,\lambda2798 \else Mg\,{\sc ii}\,$\lambda2798$ \fi}
\newcommand  \civ  {\ifmmode {\rm C}\,{\sc iv} \else C\,{\sc iv}\fi}
\newcommand  \CIV  {\ifmmode {\rm C}\,{\sc iv}\,\lambda1549 \else C\,{\sc iv}\,$\lambda1549$\fi}

\newcommand  \oi	{\ifmmode \left[{\rm O}\,{\textsc i}\right] \else [O\,{\sc i}] \fi}
\newcommand  \OI	{\ifmmode \left[{\rm O}\,{\textsc i}\right]\,\lambda6300 \else [O\,{\sc i}]$\,\lambda6300$ \fi}

\newcommand  \oii	{\ifmmode \left[{\rm O}\,{\textsc ii}\right] \else [O\,{\sc ii}] \fi}
\newcommand  \OII	{\ifmmode \left[{\rm O}\,{\textsc ii}\right]\,\lambda3727 \else [O\,{\sc ii}]\,$\lambda3727$ \fi}
\newcommand  \oiii	{\ifmmode \left[{\rm O}\,{\textsc iii}\right] \else [O\,{\sc iii}]\fi}
\newcommand  \OIII	{\ifmmode \left[{\rm O}\,{\textsc iii}\right]\,\lambda5007 \else [O\,{\sc iii}]\,$\lambda5007$\fi}

\title[Evolution of \mm]{The Evolution of \mm\ Between z=2 and z=0}

\author[Trakhtenbrot \&\, Netzer]{Benny Trakhtenbrot\thanks{E-mail: trakht@wise.tau.ac.il} and Hagai Netzer\\
School of Physics and Astronomy, Tel Aviv University, Tel Aviv 69978, Israel
}

\begin{document}

\date{Accepted 2010 May 9.  Received 2010 May 9; in original form 2010 January 25}

\pagerange{\pageref{firstpage}--\pageref{lastpage}} \pubyear{2010}

\maketitle
\label{firstpage}

\begin{abstract}
We propose a novel method to estimate \mm, the ratio of stellar mass (\mstar) to black hole mass (\mbh)
at various redshifts using two recent observational results: 
the correlation between the bolometric luminosity of active galactic nuclei 
(AGN) and the star formation rate (SFR) in their host galaxies, and the correlation between 
SFR and \mstar\ in star-forming (SF) galaxies.  
Our analysis is based on \mbh\ and \lbol\ measurements in two large samples of type-I AGN
at z$\simeq$1 and z$\simeq$2, and the measurements of \mm\ in 0.05$<$z$<$0.2 red galaxies.
We find that \mm\ depends on \mbh\ at all redshifts. 
At z$\simeq$2, \mm$\sim$280 and $\sim$40 for \mbh=$10^8$ and \mbh=$10^9$\msun, respectively. \mm\ grows by a factor of $\sim$4-8 
from z$\simeq$2 to z=0 with extreme cases
that are as large as 10--20. 
The evolution is steeper than reported in other studies, 
probably because we treat only AGN in SF hosts.
We caution that estimates of \mm\ evolution which ignore the dependence of this ratio on \mbh\
can lead to erroneous conclusions.
\end{abstract}

\begin{keywords}
galaxies: active -- galaxies: nuclei -- galaxies: evolution -- quasars: general.
\end{keywords}

\section{Introduction}
\label{sec:intro}

Study of the co-evolution of Active Galactic Nuclei (AGN) and their host galaxies provides important clues about
the growth of super massive black holes (SMBHs) and the star formation (SF) history of the Universe.
In the local universe one finds a tight correlation between the SMBH mass (\mbh) and the mass of the bulge of
its host, \mbul\ (Marconi \& Hunt, 2003; H\"{a}ring \& Rix 2004, hereafter HR04), or alternatively with the stellar velocity dispersion \sigs\ (Ferrarese \& Merrit 2000; Gebhardt \et 2000).  
Typically \mbul/\mbh$\simeq$500--1000. 
$\mbul/\mbh$ must have been smaller at high redshift. 
For example, $\mbh\sim10^{9-10}\,\Msun$ are often observed at z=3--6 (e.g. Netzer 2003; Shemmer \et 2004; Fan \et 2006), yet galaxies that are 500-1000 times more massive are never observed at z$>$0.5 and are very rare even at z$\simeq$0.

The evolution of the $\mbh-\sigs$ relationship has been studied in numerous papers.
Examples are 
Shields \et (2003) who find no evolution up to z$\simeq$2, and Woo \et (2008) who suggest that $\mbul/\mbh$ 
has increased by a factor of $\sim$3 since z=0.6. 
The uncertainties in all such measurements are very large due to the difficulties in measuring
\sigs\ or \mbul\ in high redshift AGN hosts.
While measuring \mbul\ in high-redshift galaxies is difficult, the total stellar mass, \mstar\ is easier to obtain.
This is achieved by multi-band spectral
energy distribution (SED) fitting, which is used to constrain \ml. 
However, obtaining \mm\ for AGN hosts is severely limited by the problematic subtraction of the bright point-like continuum in type-I AGN and the estimation of \mbh\ in type-II AGN.

Several recent studies used deep imaging to estimate \mm\ (or $L_{\rm Host}/M_{\rm BH}$) in z$\sim$0.5--3 AGN by resolving the
host galaxy emission and by careful PSF modelling (e.g. Kukula \et 2001; Peng \et 2006; Kotilainen \et 2007; Bennert \et 2010).
The measured host luminosity was translated to \mstar\ by \textit{assuming} a certain \ml\ ratio.
A common assumption is that AGN hosts are ``red and dead'', with a stellar population which evolves passively from a high formation redshift, e.g. $z_{\rm form}\simeq$5. 
A detailed study of this type, including a summary of many earlier findings, is given in Decarli et al. (2010; hereafter D10)
 who find that \mm\ evolves following $\mmwp\propto\,z^{-0.28}$. 
The assumption of non-SF AGN hosts may describe some objects,  
but many studies find that hosts of luminous AGN often contain
much younger stellar populations (e.g., Kauffmann \et 2003; Jahnke \et 2004; Silverman \et 2009; Merloni \et 2010, and references therein).

In this \textit{Letter} we suggest a novel method to estimate the evolution of \mm. 
Our approach makes use of the known relationships between star formation rate (SFR) and \mstar\ in SF galaxies (SFGs), 
and between the bolometric luminosity of AGN (\lbol) and the SFR in their hosts. 
We describe these relationships in \S\ref{sec:sfr_corr} and use them to estimate \mm\ in SF galaxies
 at z$\simeq$0.15, \zandz\ in \S\ref{sec:results}. 
In \S\ref{sec:discussion} we discuss the implications to the co-evolution of SMBHs and their hosts. 
Throughout this work we assume a standard $\Lambda${\sc CDM} cosmology with 
$\Omega_{\Lambda}$=0.7, $\Omega_{M}$=0.3 and $H_{0}$=70\,\kms\,Mpc$^{-1}$.

\section{SFR-\mstar\ and SFR-\lbol\ correlations}
\label{sec:sfr_corr}

Our work is based on two empirical correlations. The first is the well
established, redshift dependent, SF sequence (a correlation between SFR and total
stellar mass, \mstar) in SF galaxies. 
There are numerous papers on this issue
and the following is only a partial list which is most relevant to our work.
Brinchmann \et (2004; hereafter B04) studied a large sample of low redshift 
SFGs in the Sloan Digital Sky Survey (SDSS). 
They show a clear relationship of the form 
$SFR \simeq 8.7\,\left(\mstar/10^{11}\right)^{0.77}$.
The result was later confirmed by Salim \et (2007) who used a combination of SDSS and \textit{GALEX} data. 
Similar relationships at higher redshifts, based on mid-IR and
UV observations and multi-wavelength SED modelling, are reported in 
Elbaz \et (2007; hereafter E07), Daddi \et (2007; hereafter D07), Noeske \et (2007), Drory \& Alvarez (2008) and 
several other papers (see a more complete list in Dutton \et 2009).
E07 and D07 used the combined \textit{HST}, \textit{Spitzer} and ground-based photometry of the GOODS field.
E07 studied blue ($U-g\ltsim1.45$) galaxies at z=0.8--1.2 and found  
 $SFR\simeq57\,\left(\mstar/10^{11} \Msun\right)^{0.9}\,\mpyr$. 
D07 studied \textit{BzK}-selected galaxies at z$\simeq$1--3.
For z$\simeq$2 they find  $SFR \simeq 200\,\left(\mstar/10^{11}\right)^{0.8}\,\mpyr$.
The scatter around both relations is $\sim$0.3-0.4 dex.
Noeske \et (2007) reports $SFR\propto\mstar^{0.7}$ for z=0.2--0.7 AEGIS galaxies.
Drory and Alvarez (2008) studied the FORS Deep field up to z$\simeq$4.5 and find similar trends, albeit with systematically lower SFR.
Much of the differences between the various SFR-\mstar\ relations
can be attributed to the selective inclusion of extremely low-SFR galaxies in the samples under study.

There is mounting evidence that many, perhaps most AGN hosts are actively forming stars.
This relates to the issue of whether such hosts are ``blue'', ``red'', or  
``green valley'' sources (e.g., Brammer \et 2009 and references therein).
This is especially important at high redshift due to the known tendency for the fraction of blue galaxies to increase with redshift 
(e.g. E07 and references therein).
Works like
Salim \et (2007) show that low-z AGN hosts populate the more massive part of the SF sequence. 
These high masses, in turn, mean that the apparent ``green'' colours of many AGN hosts are the consequence of low \textit{specific} SFR (SSFR), not a low SFR. 
Moreover, Brammer et al. (2009) show that many green valley galaxies belong to the blue sequence once reddening is properly taken into account. 
While such ideas are well supported for low redshift AGN, there is a need for more evidence at high-z. 
Some such data already exists, e.g. the Silvermann \et (2009) work that claims that the SFR and \mstar\ in AGN hosts at 
z$\simeq$0.7-1 are indistinguishable from those of inactive SF galaxies. 
Given all the above, our first assumption is that the hosts of {\it most}
luminous, optically-selected AGN are part of the SF 
sequence at all redshifts. 
This assumption is further justified in \S\ref{sec_sub:res_mm_3z}.


Our second assumption is that there exists a significant correlation between the bolometric luminosity of AGN (\lbol) and 
the SFR of their hosts.  This correlation is hinted at in various intermediate and high redshift studies 
(Netzer \et 2007; Lutz \et 2008) and in low redshift type-II AGNs (Netzer 2009; hereafter N09). 
Here we adopt an updated version of the correlation presented in N09, by  
excluding LINERs from the group of low-z AGN. 
This gives 
\begin{equation}
SFR\simeq32.8\left(\lbol/10^{46}\,\ergs\right)^{0.7} \Msun\,yr^{-1}\,\,. 
\label{eq:Lbol_SFR}
\end{equation}
This equation is {\it not} a fit to the data but rather a line that goes through the points. 
As discussed in N09, there is no simple way to derive a best fit function to this inhomogeneous data set.

\begin{figure}
\centering
\includegraphics[width=7cm]{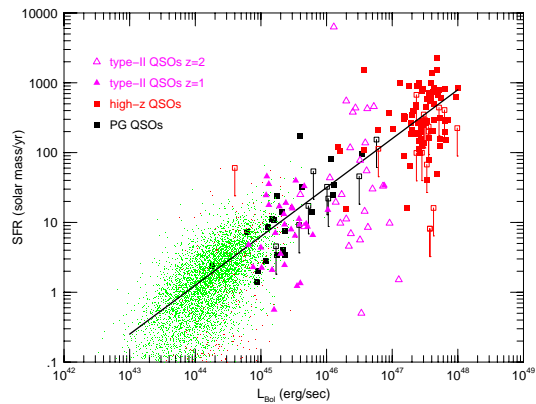}
\caption{The \lbol-SFR relation used in this work. The magenta symbols are \zandz\ zCOSMOS sources described in the text and the small
green dots are SDSS S2s. All other subsamples are described in N09. 
The solid line represents Eq.~\ref{eq:Lbol_SFR}.
}
\label{fig:Lbol_Lsf}
\end{figure}

Since the above correlation is crucial for our study,
we show the data used here in Fig.~\ref{fig:Lbol_Lsf}.
In addition to the samples presented and discussed in N09, this version also includes two additional 
groups of zCOSMOS type-II AGNs at \zandz, kindly provided  by V. Mainieri. 
These are X-ray selected sources where \lbol\ is estimated from $L(2-10\,{\rm keV})$
and the SFR is based on multi-wavelength SED fitting.
The two groups lie on the above relationship with a scatter in SFR of about 0.4 dex, comparable to the 
overall scatter. There are several reasons for this scatter.   
At the low-\lbol\ end, the scatter is related to the inaccuracies in SFR and \lbol\ determination as 
well as real scatter in these properties. 
At \zandz, the scatter reflects the uncertainty in SED modelling
and the range of \mstar\ across the SF sequence.
At the high-\lbol\ end, much of the scatter is due to observational 
uncertainties, the incompleteness of the high redshift AGN samples 
and, perhaps, the extreme SFR in mergers.
A more complete account of these issues can be found in N09.

The following analysis relies on the combination of the above correlations for SF AGN hosts. 
Given \lbol, we can determine SFR (Eq.~\ref{eq:Lbol_SFR}) and this, in turn, 
can be translated to \mstar\ given the redshift dependent SFR-\mstar\ correlations.

\section{The redshift evolution of \mstar/\mbh}
\label{sec:results}

\subsection{\mm\ at low and high redshift}
\label{sec_sub:res_mm_3z}

To explore \mm, we define several samples of both AGN and non-AGN galaxies.
The first is a large sample of 0.05$<$z$<$0.2 galaxies from the value-added MPA/JHU SDSS DR4 database.
This includes photometry, \mstar, SFR and \sigs\ for all sources (B04).
To justify the use of the \mbh-\sigs\ relation of Tremaine \et (2002), we choose only red galaxies, following the colour cut of Baldry \et (2004).
The number of such sources, after applying some basic quality criteria, is 210,158.  
They cover the range $5\times10^9<\mstar/\msun\ltsim6\times10^{11}$ and $10^6<\mbh/\msun\ltsim2\times10^9$ (the latter lower limit was chosen to filter out dubious measurements).
\mstar\ is derived from the SDSS photometry and thus the flux limit of the sample
 ($r_{\rm petro}<17.7$) affects the number of low-\mstar\ galaxies. Here we deal mostly with the larger \mstar\ 
systems and these limitations do not affect our general conclusions.
 
The work involves also four AGN samples: type-I and type-II samples at 0.1$<$z$<$0.2, a type-I sample at
z$\simeq$1 and a type-I sample at z$\simeq$2. The first two provide little new information but serve to test and
to justify the general new method presented here. The 0.1$<$z$<$0.2 redshift range is chosen to enable proper measurements of
type-I and type-II AGN (see details in N09).
The type-II AGNs are ``strong AGN'' (Seyfert 2s - S2s but not LINERs)
drawn from the local SDSS sample.
\lbol\ for these 1152 AGN is determined by the \oi\ and \oiii\ method of N09.
The 2814 type-I AGN are drawn directly from the SDSS/DR7 (Abazajian \et 2009) database and are analyzed in a way similar to the one presented in Netzer \& Trakhtenbrot (2007; hereafter NT07). 
Here \mbh\ is estimated from the monochromatic luminosity at 5100\AA\ (\Lop) and the FWHM of the \hb line.
\lbol\ is determined from \Lop, using the bolometric correction factors of Marconi \et (2004).
To avoid selection biases, we applied a common \textit{observed} flux limit for the type-I and type-II samples, such that only 
sources with F(\oiii)$>4\times10^{-16}\,\ergcms$ are included. 
Assuming a typical bolometric correction of 2500 (Netzer \et 2006, NT07), this translates to  $\lbol>2.5\times10^{43}\,\ergs$.

We constructed the SF sequence for our type-II low redshift sample and compared it with 
the SF sequence presented in B04. The agreement is very good, with about 70\% of the AGN hosts lying on the sequence.
We also confirm earlier results (e.g. Salim \et 2007) 
which suggest that AGN hosts concentrate at somewhat higher \mstar.
We find that despite their apparently red colours, the majority of the red S2 hosts are actively forming stars and are
situated on the SF sequence.
These SF S2 hosts are, in fact, mostly green valley sources, with lower SSFR.
Some 30\% of the AGN are not part of the SF sequence. 
We suspect that the fraction of such sources at higher redshift is smaller but have no way to check it qualitatively. 
The remaining of the paper and the results concerning \mm\ refer {\it only} to those AGN that are on the SF sequence.

Next, we translate \lbol\ to SFR for the 0.1$<$z$<$0.2 type-I AGN using Eq.~\ref{eq:Lbol_SFR}.
This is then converted to \mstar\ using the low redshift SF sequence of B04. 
Combining with the measured \mbh, we obtain \mm\ for all these sources. 
This can be compared with \mm\ measured directly for the type-II z=0.1--0.2 AGN.  
The overall good agreement justifies the use of a similar procedure for higher redshift AGN samples.

We selected two higher redshift type-I AGN samples from the SDSS/DR7.
We chose sources in the range 0.9$<$z$<$1.1 and 1.8$<$z$<$2 (5330 and 4352 objects, respectively), which are the nominal redshift ranges of the SFG  samples of E07 and D07.  
Full details of the sample selection, line fitting and related analysis will be given in a forthcoming 
publication.
In short, we use a similar procedure to the one described in NT07 and in Shen \et (2008) to measure 
the \MgII emission line complex.
The FWHM of the line and the adjacent continuum luminosity (\Lthree) are combined to estimate \mbh\ 
using the relation of McLure \& Dunlop (2004). 
We calculate \lbol\ from \Lthree\ by calibrating \Lthree\ against \Lop\ in a separate sub-sample where the two continuum
bands are observed in the spectrum. 
We then estimated the SFR by using Eq.~\ref{eq:Lbol_SFR}.
The distribution of inferred SFRs (not shown here) clearly shows that these AGN hosts are actively forming stars at rates that are comparable to the non-AGN SFGs at those redshifts. 
In particular, 99\% of the z$\simeq$2 sources have SFR$\gtsim42\mpyr$, the median SFR in the SF sequence of D07.

While most (not all, see earlier comments) high redshift AGN hosts are expected to lie on the SF sequence, the range in SFR and in \mstar\ can be very different from those found for non-AGN samples because of the differences in properties of the observed samples, in particular different flux limits. 
To examine this in detail, we focus on our z$\simeq$2 type-I AGN sample. 
We compare the range of derived SFR and possible range of \mstar\ to the same properties in the D07 sample, using data kindly provided by Emanuele Daddi.
Such a comparison involves two crucial factors. 
First, the D07 sample includes much fainter sources.
As explained above, almost all our type-I AGN occupy only the upper part of the z$\simeq$2 SF sequence. 
This is a direct consequence of the SDSS flux limit. 
Second, one can consider two approaches to deduce typical values of \mstar\ for the \zandz\ samples, by either 
(1) converting each individual SFR to \mstar\ through the best fit SF sequence, at the appropriate redshift (i.e. E07 and D07), or 
(2) sampling the distribution of \mstar\ per given SFR (in the high-redshift SF sequences), for each deduced value of SFR. 
By definition, the latter will result in a considerably broader distribution of the derived \mm, due to the wide range of properties in the observed samples. 
In the following analysis we only use the first approach, i.e. we derive \mstar\ for each \zandz\ AGN host, by converting its (derived) SFR through the E07 and D07 relations, respectively.
We note that the $\sim$0.4 dex scatter in these relations is a real uncertainty on our high-redshift results.
The z$\simeq$2 sample covers $9.5\times10^7\ltsim\mbh/\msun\ltsim5\times10^9$ and $1.7\times10^{10}\ltsim\mstar/\msun\ltsim1.8\times10^{11}$, 
while the z$\simeq$1 sample covers $3.8\times10^7\ltsim\mbh/\msun\ltsim2.5\times10^9$ and $2.8\times10^{10}\ltsim\mstar/\msun\ltsim2.9\times10^{11}$.
Clearly, the varying ranges of \mstar\ and \mbh\ in the different local and high-redshift samples prohibit a simplistic comparison of the mean \mm. 
In what follows we thus preform a more careful comparison.

\vspace*{-0.4cm}

\subsection{\mm\, evolution}
\label{sec_sub:res_mm_evo}

\begin{figure}
\centering
\includegraphics[width=6.2cm]{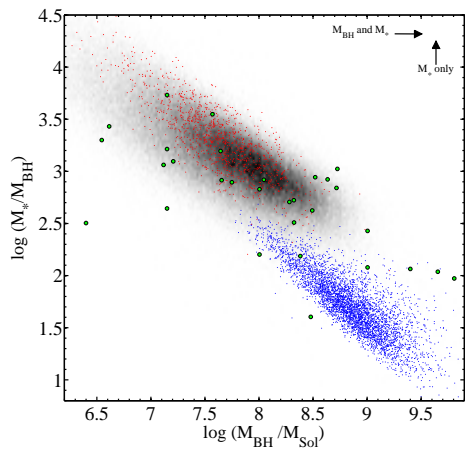}
\caption{
\mm\ for different local and high redshift samples: 
local red galaxies (gray scale density map), type-II AGN (red points), the HR04 sources (green circles) and z$\simeq$2 type-I AGN (blue points).
The arrows demonstrate simple scenarios where only \mstar\ or both \mstar\ and \mbh\ grow by a factor of 2.
}
\label{fig:MM_vs_Mbh_all}
\end{figure}

Standard galaxy evolution scenarios suggest that the end phase of many high redshift SF galaxies are massive, red ellipticals.
Therefore, we use our data to compare the
properties of the \zandz\ AGN hosts to those of red galaxies in the local Universe. 
Fig.~\ref{fig:MM_vs_Mbh_all} shows the entire sample of 0.05$<$z$<$0.2 red galaxies, as a gray scale density map. 
The galaxies form a well-defined band in the \mm-\mbh\ plane that follows the approximate relationship
 \mm$\propto$\mbh$^{-0.7\pm0.1}$.
Also shown is the sample of 0.1$<$z$<$0.2 type-II AGN that follows a similar trend.

Fig.~\ref{fig:MM_vs_Mbh_all} also shows data for the 30 local galaxies from HR04 with dynamically measured \mbh. 
In this case we plot \mbul\ rather than \mstar. 
All \mbh\ are taken directly from HR04
except  those of M\,87, NGC\,4649 and NGC\,4697 where we used new
 measurements reported in Gebhardt \& Thomas (2009), Shen \& Gebhardt (2010) and Forestell \et (2010). 
This sequence extends up to $\mbh\simeq6\times10^9\msun$ and 
\mm$\sim$100 and
shows a clear dependence on \mbh, similar
to the red galaxies in our sample and in contradiction to
to claims of a constant \mm$\sim$500--1000.
Several of the HR04 galaxies are, in fact, disk+bulge systems with relatively small bulges 
and the replacement of \mbul\ by \mstar\ will enlarge the plotted \mm. 
Based on this fact and the large uncertainties on the estimated \mm,
we suggest that our sample of local red galaxies is consistent with the HR04 sequence.
We also show in Fig.~\ref{fig:MM_vs_Mbh_all} \mm\ and \mbh\ for our z$\simeq$2 sample. 
The high redshift objects form a similar sequence, shifted down by about 0.75 dex,
with approximately the same slope.
The z$\simeq$1 sequence (not shown) is situated between the two.

Fig.~\ref{fig:MM_vs_Mbh_all} provides important clues about the redshift evolution of \mm. 
First, \mm\ evolution must be considered for a given \mbh. 
The population mean of \mm\ would fail to account for the fact that the way such samples are drawn 
is biased towards larger \mbh\ at higher redshifts. 
To demonstrate this, we show in Fig.~\ref{fig:MM_hist_3z_3m} three histograms, representing three vertical cuts in Fig.~\ref{fig:MM_vs_Mbh_all} for 0.2 dex wide \mbh\ bins centred at $10^{8}$, $10^{8.5}$ and $10^{9}$ \msun. These correspond to the case of \mstar\ {\it but no} \mbh\ growth.
The typical growth factors of \mm\ between z=2 and z$\simeq$0.1 are
$\sim$3.8 for \mbh=$10^{8}$\msun,
$\sim$5.7 for \mbh=$10^{8.5}$\msun, and 
$\sim$7.8 for \mbh=$10^{9}$\msun.
We also show, in dashed vertical lines, the mean \mm\ in the HR04 sample for the same \mbh\ 
obtained from the best ({\sc BCES} bisector) linear fit to their data.
The corresponding \mm\ growth factors are  
$\sim$2, $\sim$3.3, and $\sim$5.6, for the same values of \mbh.

\vspace*{-0.4cm}

\section{Discussion}
\label{sec:discussion}

The analysis presented here suggests strong evolution in \mm\ up to z=2, 
much steeper than what is suggested in other studies. 
In particular, our mean values of \mm\ at \zandz\ are smaller than the ones reported in D10,
which is the most up-to-date compilation of such works. 
There are three main reasons for these differences:
\begin{figure}
\centering
\includegraphics[width=6.2cm]{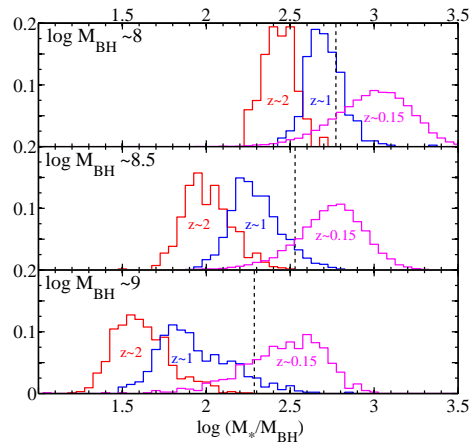}
\caption{The distributions of \mm\ at different redshift, for the three \mbh\ sub-groups discussed in the text.}
\label{fig:MM_hist_3z_3m}
\end{figure}
\begin{enumerate}
 \item 
As already mentioned, most earlier works estimate \mstar\ by assuming a passive evolution from a high formation redshift.
This is obviously an over-simplification of the evolution of most galaxies  
and significant epochs of SF at z$<$4 
(e.g. D07; E07; Drory \& Alvarez 2008; van Dokkum \et 2010, and references therein). 
Accounting for younger stellar populations would result in lower \mstar\ and thus lower \mm\ (see D10 and Peng \et 2006). 
\item
The D10 sample represents the \textit{minority} of the AGN population, as hinted by two biases.
First, the majority of the D10 sources lie in the top 15\% of the \mbh\ distributions corresponding to their redshift. 
Second, the selection criteria for the D10 HST observations could have been biased towards large, resolved galaxies with large \mstar.
\item
The mean \mbh\ in high redshift AGN samples is systematically larger than the corresponding low redshift \mbh.
For example, in our large sample of red galaxies, 99\% of the sources show \mbh$<10^{8.8}$\msun\
 while 20\% of the z$\simeq$1 and 57\% of the z$\simeq$2 AGN have larger \mbh. 
All the z$\simeq$2 sources in D10 have \mbh$\gtsim10^{8.7}$\msun. 
Such objects should only be compared with local galaxies which host BHs that are at least as massive. 
As Fig.~\ref{fig:MM_vs_Mbh_all} shows, this corresponds to \mm=100--200, 
instead of the commonly used $\sim$700.
\end{enumerate}
In conclusion, while our work applies to most AGN, the D10 sample probably represents the remaining sources.

The results presented here point to a scenario 
where many galaxies have to increase their mass by factors of 4-8 (2-4) 
since z$\simeq$2 (z$\simeq$1). 
The growth factors for the most massive BHs are not well determined  
since the number of very massive galaxies in the
local universe is not large enough to reliably extend the results of Fig.~\ref{fig:MM_vs_Mbh_all} beyond $\mbh\simeq10^9\msun$.
These numbers represent the requirement for the galaxies to \textit{over}-grow their SMBHs by the above factors.
This seems to be consistent with models which suggest that the high mass SMBHs observed at z$\simeq2$ could have accumulated 
most of their mass by that redshift (Marconi \et 2004).
On the other hand, it may be in contradiction with at least some scenarios linking AGN activity to the shut-down of SF in their host galaxies (see Somerville \et 2008; Cattaneo \et 2009 and references therein). 

While a full discussion of the various growth scenarios of \mstar\ is beyond the scope of this \textit{Letter}, we comment briefly on some of these ideas.
Major galaxy mergers would increase both \mstar\ and \mbh, either through starbursts and gas accretion in ``wet mergers'' or the possible coalescence of the two SMBHs involved in ``dry mergers''.
However, theoretical and observational arguments
(e.g. Lotz \et 2008; Genel \et 2009) suggest a low rate of such events for z$<2$ galaxies.
Thus, major mergers cannot change \mm\ by more than a factor of $\sim2-3$ between z=2
and z=0.
Small  ``dry mergers'' may help. 
For example, Naab, Johansson \& Ostriker (2009) show that present-day massive red ellipticals gain 
$\sim40\%$ of their mass through accretion of smaller companions since z$\sim2$.
Intense SF in outer parts of galaxies due to external source of cold gas
which does not find its way to the centre (e.g. van Dokkum \et 2010), is another possibility.
More possibilities and more references are discussed in Benson \& Devereux (2009).
The \mm\ distributions presented here suggest an increase in \mm\ by factors
beyond what is suggested in many theoretical studies.

Finally, we comment on the possibility that the suggested evolution of \mm\ 
could be due to two wrong assumptions. 
First, many more AGN hosts may not lie on the SF sequence or may not obey the \lbol-SFR correlation used here. 
This is unlikely to be the case at low redshift, where SDSS type-II AGNs are used. 
However, the selection of at least some of the most luminous, high redshift  
high-\lbol\ sources in Fig.~\ref{fig:Lbol_Lsf} may be biased towards high FIR luminosity, high SFR AGN 
hosts (e.g. Zheng \et 2009) in particular if these are found in mergers that are not part of the SF sequence.  
\textit{Herschel} observations of well-defined AGN samples are likely to resolve this issue.
Second, the Drory \& Alvarez (2008) work shows a decline in SFR at the high-\mstar\ end for
 z$<$2 galaxies. 
Thus, some of our \lbol-based SFR estimates might be associated with considerably larger values of \mstar. 
The E07 and D07 SF sequences do not show such a decline.

We conclude that there is a steep evolution in \mm\ from z$\simeq2$ to z=0 for SF AGN hosts.
This trend is barely consistent with some, but not all galaxy and BH evolution models.
We have also demonstrated the crucial importance of considering different \mbh\ groups separately when evaluating the \mm\ evolution.

\vspace*{0.4cm}

\noindent
We thank an anonymous referee for very useful comments on the manuscript. 
We thank Ido Finkelmann for fruitful discussions;  
Vincenzo Mainieri for allowing us to use zCOSMOS results ahead of publication; 
 Roberto Decarli and Emanuele Daddi for sending us their data; 
Niv Drory and Samir Salim for their thoughtful comments; and the MPA/JHU team for making their SDSS catalogues available to the public. 
This study makes use of data from the SDSS  
(http://www.sdss.org/collaboration/credits.html).
Funding for this work has been provided by the Israel Science
Foundation grant 364/07.

\vspace*{ -0.6cm}

{}

\end{document}